\documentstyle[12pt]{article}
\begin{document}

\begin{center}
{\bf HEAVY QUARK HADROPRODUCTION IN $k_T$-FACTORIZATION APPROACH
AND THE EXPERIMENTAL DATA}

\vspace{0.5cm}

Yu.M.Shabelski and A.G.Shuvaev \\
Petersburg Nuclear Physics Institute, \\
Gatchina, St.Petersburg 188350 Russia \\

\end{center}

\vspace{0.5cm}

\begin{abstract}
We compare the numerical predictions of the $k_T$-factorization approach
(semi\-hard theory) for heavy quark production in high energy
nucleon-nucleon and photon-nucleon collisions with the experimental
data from Tevatron-collider and HERA. Predictions for heavy quark
production at Tevatron, LHC and HERA are also presented.

\end{abstract}

\vspace{2cm}

E-mail SHABELSK@THD.PNPI.SPB.RU

E-mail SHUVAEV@THD.PNPI.SPB.RU

\newpage

\section{Introduction}

The investigation of heavy quark production in high energy hadron
collisions is an important method for studying the quark-gluon
structure of hadrons. The description of hard interactions in hadron
collisions within the framework of QCD is possible only with the help of
some phenomenology, which reduces the hadron-hadron interaction to the
parton-parton one via the formalism of the hadron structure functions.
The cross sections of hard processes in hadron-hadron interactions can
be written as the convolutions of sub-process QCD matrix elements
squared with the parton
distributions in the colliding hadrons.

The most popular and technically simplest approach is the so-called QCD
collinear approximation, or parton model (PM). In this model all
particles involved are assumed to be on mass shell, carrying only
longitudinal momenta, and the cross section is averaged over two
transverse polarizations of the incident gluons. The virtualities $q^2$
of the initial partons are taken into account only through their
structure functions. The cross sections of QCD subprocess are calculated
usually in the next to leading order (NLO) \cite{1,2,NDE,Beer,Beer1} of
$\alpha_S$ expansion. The transverse momenta of the incident partons are
neglected in the QCD matrix elements. This is the direct analogy of the
Weizsaecker-Williams approximation in QED.

Another possibility to incorporate the incident parton transverse
momenta is referred to as $k_T$-factorization approach
\cite{CCH,CE,MW,CH,CC,HKSST}, or the theory of semihard interactions
\cite{GLR,LR,8,lrs,3,Zo1,SS,BS,Zo2}.  Here the Feynman diagrams are
calculated with account of the virtualities and possible
polarizations of the incident partons. In the small $x$ domain there
are no grounds to neglect the transverse momenta of the gluons,
$q_{1T}$ and $q_{2T}$, in comparison with the quark mass and
transverse momenta, $p_{iT}$.  Moreover, at the very high energies and
$p_{iT}$ the main contribution to the cross sections comes
from the region $q_{1T} \sim p_{1T}$ or $q_{2T} \sim p_{1T}$, see
\cite{our,our1,our2} for details. The QCD matrix elements of the
sub-processes are rather complicated in such an approach. We have
calculated them in the LO. On the other hand, the multiple emission of
soft gluons is effectively included here. That is why it is the question
which approach is more constructive.

The $k_T$-factorization approach is based on so-called unintegrated parton
distributions which, at the moment, are known with insufficient accuracy.
Here we find them with the help of the realistic gluon
distribution GRV94 \cite{GRV} which is compatible with the most recent
data, see discussion in Ref.~\cite{GRV1}.

In Sect.~2 we present very shortly the formalism of the
$k_T$-factorization approach. The comparison of our numerical results
with some experimental data on beauty production at Tevatron-collider
together with the predictions including LHC energy are discussed in
Sect.~3. Predictions of the $k_T$-factorization approach for heavy
quark photoproduction for HERA are given in Sect.~4.

\section{Heavy quark production in the $k_T$-factorization approach}

The main contribution to the cross section of heavy quark production
at small $x$ is known to come from $gg \to Q\bar{Q}$ subprocess (and
$\gamma g \to Q\bar{Q}$ in the case of photoproduction). The transverse
momenta of the incident gluons in the small-$x$ region result in the
$k_T$-factorization approach from $\alpha_S \ln k_T^2$ gluon diffusion.
This is described by the so-called unintegrating gluon distribution
which can be written in the form \cite{our1,MW1,KMR}
\begin{equation}
f_g(x,q_T,\mu) = \sum_{a'}\left[\frac{\alpha_s}{2 \pi}
\int^{\Delta}_x P_{ga'} (z) a'\left(\frac xz,q_T^2 \right) dz
\right] T_g(q_T,\mu) \;,
\end{equation}
where the survival probability $T_g$ that gluon $g$ with the transverse
momentum $q_T$ remains untouched in the evolution up to the
factorization scale $\mu$ is
\begin{equation}
\label{Sud}
T_g(q_T,\mu) = \exp \left[ -\int^{\mu^2}_{q^2_T}
\frac{\alpha_s(p_T)}{2\pi} \frac{dp^2_T}{p^2_T}
\sum_{a'} \int^{\Delta}_0 P_{a'g} (z') dz' \right] \;.
\end{equation}
Here $a'(x,q_T^2)$ denotes $xg(x,q_T^2)$ or $xq(x,q_T^2)$, and
$P_{ga'}(z)$ are the splitting functions.

We have to emphasize that $f_g(x,q_T,\mu)$ is just the quantity which
enters into the Feynman diagrams. The distributions
$f_g(x,q_T,\mu)$ involve two hard scales\footnote{This property is
hidden in the conventional parton distributions as $q_T$ is
integrated up to the scale $\mu$.}: $q_T$ and the scale $\mu$ of the
probe. The scale $\mu$ plays a dual role. On the one hand it acts as
the factorization scale, while on the other hand it controls the
angular ordering of the partons emitted in the evolution
\cite{MCi,CFM,Mar}. The factorization scale $\mu$ separates the
partons associated with the emission off
both the beam and target protons (in $pp$ collisions) and
off the hard subprocess. For example, it separates emissions off
the beam (with polar angle $\theta < 90^o$ in c.s.m.) from those off
the target (with $\theta > 90^o$ in c.s.m.), and from the
intermediate partons from the hard subprocess. This separation was
proved in \cite{MCi,CFM,Mar} and originates from the destructive
interference of the different emission amplitudes (Feynman diagrams)
in the angular boundary regions.

If the longitudinal momentum fraction is fixed by the hard
subprocess, then the limits of the angles can be expresseed in terms
of the factorization scale $\mu$ which corresponds to the upper limit
of the allowed value of the $s$-channel parton $k_T$.


The expression (1) with the survival probability (2) provides the
positivity of the unintegrated probability $f_g(x,q_T,\mu)$ in the
whole interval $0 < x < 1$.

Here we deal with the matrix element accounting for the gluon
virtualities and polarizations. Since it is much more complicated than
in the PM we consider only the LO of the subprocess $gg \to Q\bar{Q}$
which gives the main contribution to the heavy quark production cross
section at small $x$, see the diagrams a, b and c in Fig.~1.  The lower
and upper ladder blobs present the unintegrating gluon distributions
$f_g(y,q_{1T},\mu)$ and $f_g(x,q_{2T},\mu)$.

The differential cross section of heavy quark hadroproduction has the
following form:\footnote{We put the argument of $\alpha_S$ to be equal
to gluon
virtuality, which is very close to the BLM scheme\cite{blm}; (see also
\cite{lrs}).}
\begin{eqnarray}
\frac{d\sigma_{pp}}{dy^*_1 dy^*_2 d^2 p_{1T}d^2
p_{2T}}\,&=&\,\frac{1}{8\pi^2}
\frac{1}{(s)^2}\int\,d^2 q_{1T} d^2 q_{2T} \delta (q_{1T} +
q_{2T} - p_{1T} - p_{2T}) \nonumber \\
\label{spp}
&\times &\,\frac{\alpha_s(q^2_1)}{q_1^2} \frac{\alpha_s (q^2_2)}{q^2_2}
f_g(y,q_{1T},\mu) f_g(x,q_{2T,},\mu) \vert M_{QQ}\vert^2.
\end{eqnarray}
Here $s = 2p_A p_B\,\,$, $q_{1,2T}$ are the gluon transverse momenta
and $y^*_{1,2}$  are the heavy quark rapidities in the hadron-hadron
c.m.s. frame,
\begin{equation}
\label{xy}
\begin{array}{lcl}
x_1=\,\frac{m_{1T}}{\sqrt{s}}\, e^{-y^*_1}, &
x_2=\,\frac{m_{2T}}{\sqrt{s}}\, e^{-y^*_2},  &  x=x_1 + x_2\\
y_1=\, \frac{m_{1T}}{\sqrt{s}}\, e^{y^*_1}, &  y_2 =
\frac{m_{2T}}{\sqrt{s}}\, e^{y^*_2},  &  y=y_1 + y_2 \\
&m_{1,2T}^2 = m_Q^2 + p_{1,2T}^2. &
\end{array}
\end{equation}
$\vert M_{QQ}\vert^2$ is the square of the matrix element for the heavy
quark pair hadropro\-duction. It is calculated in the Born approximation
of QCD without standard simplifications of the parton model. Contrary to
what is mentioned in \cite{BS}, the transformation Jacobian from $x, y$ to
$y^*_1, y^*_2$ is included in our matrix element.

We take
\begin{equation}
m_c = 1.4\; {\rm GeV}, \qquad m_b = 4.6\; {\rm GeV} \;,
\end{equation}
for the values of short-distance perturbative quark masses
\cite{Nar,BBB}.

\section{Beauty production at Tevatron-collider}

Eq.~(3) enables us to calculate straightforwardly all one-particle
distributions as well as correlations between two produced heavy quarks
using, say, the VEGAS code \cite{Lep}.

However there exists a principle problem coming from the infrared
region.  Since the functions $f_g(y,q_{1T},\mu)$ and $f_g(x,q_{2T},\mu)$
are unknown at small values of $q^2_2$ and $q^2_1$, i.e.
in nonperturbative domain we calculate separately the contributions
from $q^2_1 < Q^2_0$, $q^2_2 < Q^2_0$, $q^2_1 > Q^2_0$ and
$q^2_1 > Q^2_0$ \cite{our,our1,our2}.

The first contribution ($q^2_1 < Q^2_0$, $q^2_2 < Q^2_0$) with the
matrix element averaged over directions of the two-dimensional vectors
$q_{1T}$ and $q_{2T}$ is exactly the same as the conventional LO parton
model expression. It is multiplied by the 'survival' probability
$T^2(Q_0^2,\mu^2)$. (We assume $Q_0^2$ = 1~GeV$^2$ in the following
numerical calculations.) In this contribution the sum of the produced
heavy quark transverse momenta is taken to be  exactly zero.

The next three contributions (from the domains where one, or both $q^2_i >
Q_0^2$) contain the corrections to the parton model matrix element due to
gluon polarizations, virtualities and transverse momenta. Their relative
contribution strongly depends on the initial energy. If it is not high
enough, the first term dominates, and all results are similar to the
conventional LO parton model predictions \cite{our1}. In the case of very
high energy the opposite situation takes place, the first term can be
considered as a small correction and our results differ from the
conventional ones. So the highest energies are needed to observe (and to
study) the $k_T$-factorization effects.

In the numerical calculations some uncertainties come from the value of
cut-off $\Delta$ in Eqs.~(1), (2). In particular, in \cite{our1} we
used
\begin{equation}
\Delta = 1 - q_T/\mu \;.
\end{equation}
The angular ordering \cite{MCi,CFM,Mar} implies that the cut-off
\begin{equation}
\Delta = \frac{\mu}{\mu +q_T} \;.
\end{equation}
In this case one gets the non-zero values of $f_a(x, q_T, \mu)$ even at
the large $q_T > \mu$. This is especially important at the high energies
where the essential values of $x$ and $z$ in Eqs.~(1), (2) are very
small.

From the formal point of view the difference between (6) and (7) is
beyond the DGLAP LO accuracy. With the same accuracy we can suggest
\begin{equation}
\Delta = \frac{\mu}{\sqrt{\mu^2 +q_T^2}} \;,
\end{equation}
or something like that.

Another source of uncertainties comes from the fact that the
unintegrated gluon distributions are not known with the needed accuracy
(from the evolution equation and the experimental data).

Let us compare the results of our numerical calculations with the data
on beauty production at Tevatron-collider. In Fig. 2a we present the
data \cite{Abb} on $b\bar{b}$ pair production with $p_T > p_{min}$
identified by their muon decays, as the function of $p_{min}$ at
$\sqrt{s}$ = 1.8~TeV. The curves of different type show the calculated
results with different cut-off $\Delta$ in (1) and (2), whereas the upper
and lower curves of the same type are obtained with different QCD scales,
$\mu^2 = \hat{s}$ and $\mu^2 = \hat{s}/4$, where $\hat{s}$ is the invariant
energy of the produced heavy quark pair, $\hat{s} = xys$.

One can see that the scale dependence is negligible here. The
dependence on the cut-off $\Delta$ is more important. The expressions (6)
(dash-dotted curves) and (7) (dashed curves) give too small cross
section, whereas the cut-off (8) (dotted curves) overestimates the data.
We obtain the best agreement with the experiment (solid curves),
assuming the cut-off smaller than (8) and larger than (7), namely
\begin{equation}
\Delta = \frac{\mu}{\sqrt{\mu^2 + q_T^2 +\mu q_T}}  \;.
\end{equation}

The same results for inclusive $b$-quark production at large $p_T$
extracted from muon taggeed jets \cite{Abb1} are presented in Fig. 2b.
Here again the data are in reasonable agreement with our calculations
using Eq. (9).

The rapidity distributions of the produced beauty \cite{Abb2} are shown
in Fig.~3a, b. They decrease more steeper than our predictions, but the
numerical difference is not large.

The correlation between the produced heavy quark and antiquark
\cite{HKSST,Abe} which are more sensitive to the dynamics of the
production mechanism are presented in Fig.~4. Here the transverse momentum
of one beauty quark, $p_{1T}$, is taken more than some value (6.5 GeV in
Fig.~4a and 8.75 GeV in Fig.~4b), and cross sections of second quark
production with $p_{2T} > p_{min}$ are presented as the functions of
$p_{min}$. Again, these data can be described using the cut-off (9).

There are the correlations which are more sensitive to the QCD scale
value $\mu$ in comparison with the cut-off value. Some examples are
presented in Fig.~5, where the distributions over the pair transverse
momenta are plotted with the condition that one heavy quark has fixed
(and large) transverse momentum. In all cases the shapes of the curves
calculated with $\mu^2 = \hat{s}$ and $\mu^2 = \hat{s}/4$ are
qualitatively different.

The results for the total cross sections of heavy quark production at
FNAL and LHC are presented in the Table. These cross sections are
essentially larger in comparison with the conventional NLO QCD
predictions, see, e.g. \cite{Nas}. It can be important for the
planning of the experiments at LHC.

\vskip 10 pt
\begin{center}
The total cross sections of charm and beauty production calculated in
the $k_T$-factorization approach with cut-off (9) and $\mu^2 = \hat{s}$

\vskip 20 pt
\begin{tabular}{|c|r|r|r|r|} \hline

& \multicolumn{2}{c|}{all rapidities} &
\multicolumn{2}{c|}{$\vert y_1^*\vert < 1, \vert y_2^*\vert < 1 $}
\\ \hline

$\sqrt{s}$ & $c\bar{c}$ & $b\bar{b}$ & $c\bar{c}$ & $b\bar{b}$
\\ \hline

14 TeV   & 20.9 mb & 1.16 mb & 2 mb & 142 $\mu$b  \\ \hline

1.8 TeV  & 3.15 mb & 107 $\mu$b & 423 $\mu$b & 20 $\mu$b  \\ \hline
\end{tabular}
\end{center}
\vskip 10 pt

The azimuthal correlations of heavy quarks, i.e. the distribution over
the opening angle $\phi$ between the momentum vectors of the produced
quarks in the plane transverse to the beam axis, are sensitive to the
details of the production mechanism. For example, the experimental data
at fixed target energies are in contradiction with NLO QCD collinear
approximation \cite{BEAT,BEAT1}. Our predictions for $b\bar{b}$
production at 1.8~TeV and 14~TeV are presented in Fig.~6. The same
predictions obtained for the large transverse momenta of
both quarks, $p_{1,2T} >$6~GeV, are also shown.

\section{Heavy quark photoproduction}

In the case of high energy heavy quark photoproduction we should
consider (in LO QCD) two possibilities, direct production in the
photon-gluon fusion, $\gamma p \to Q\bar{Q}$, Fig.~7, and the
resolved production via quark-gluon structure of the photon. In the
latter case the diagrams are similar to those in Fig.~1, where one
proton should be replaced by photon.

The cross section of the direct interaction can be written as
\begin{eqnarray}
\frac{d\sigma_{\gamma p}}{d^2p_{1T}} & = & \frac{\alpha_{em}e^2_Q}{\pi}
\int dz d^2q_T \frac{f_g(x,q_T,\mu)}{q^4_T} \alpha_s(q^2)
\nonumber \\ & \times &
\left\{ [(1-z)^2+z^2] \left(\frac{\vec{p}_{1T}}{D_1} +
\frac{\vec{q}_T-\vec{p}_{1t}}{D_2}
\right )^2 + m^2_Q \left(\frac1{D_1} - \frac1{D_2} \right )^2 \right\}
\;,
\end{eqnarray}
\begin{equation}
D_1 = p^2_{1T} + m^2_Q \;,\;\;
D_2 = (\vec{q}_T - \vec{p}_{1T})^2 + m^2_Q \;,
\end{equation}
where $\alpha_{em} = 1/137$ and $e_Q$ is the electric charge of the
produced heavy quark.

The resolved contribution is determined by Eq.~(3), where one of the
structure functions $f_g$ describes the partons distribution in
the proton and the other stands for the parton distribution in the photon.

The energy dependences of the total cross sections of $c\bar{c}$ and
$b\bar{b}$ photoproduction are presented in Fig.~8. The predictions for
the cross sections of charm photoproduction together with the data
\cite{ZEUS,H1a} are shown in Fig.~8a. The averaged cross section value
at low energies is taken from \cite{H1a}. The data on beauty
photoproduction cite{H1} are shown in Fig.~8b. The cut-off value (9) is
again in reasonable agreement with the data. The values (6) and (7) give
too small cross section, as well as NLO QCD approach \cite{H1}. These
results are in agreement with \cite{Zo3}, where different assumptions were
used.

We can see that the cut-off expression (9) gives a better agreement with
all the presented data in comparison with the other expressions (6)-(8).
The $\mu^2$ dependence of all distributions is rather week except of the
special case similar to Fig.~5. So, as a rule, we will present below the
results obtained with the cut-off (9) and $\mu^2 = \hat{s}$ only.

The one-particle $p_T$-distributions, $d \sigma /dp_T$, calculated
in the $k_T$-factorization approach for HERA energy, are presented in
Fig.~9. Here we give separately the sum of direct and resolved
contributions, as well as the only resolved ones. The resolved
contributions decrease with $p_T$ more fast, that is connected with the
finite phase space value. At very high energies the resolved
contributions will decrease more slow.

The predicted rapidity distributions, $d \sigma /dy^*$, of the
produced heavy quarks are shown in Fig.~10 for all events and for
events with $p_T >$ 6 GeV. The direct production gives the narrow peak
in the photon fragmentation region. This peak moves to the central
region when only quarks with high transverse momenta are registrated,
that is in agreement with the data \cite{ZEUSa}. The resolved photon
contribution dominates in the target fragmentation and partly in the
central regions.

The distributions of pair transverse momenta with the condition
that one heavy quark has fixed (and large in comparison with its mass)
transverse momentum are given in Fig.~11. Here again, as in Fig.~5, the
shapes of the curves calculated with $\mu^2 = \hat{s}$ and
$\mu^2 = \hat{s}/4$ are qualitatively different that allows one to
determine the scale value.

The predictions for the azimuthal correlations in heavy quark
photoproduction at HERA energy are presented in Fig.~12. Their
dependences on $\phi$ and on the restriction of the minimal
transverse momenta of the produced quarks are qualitatively similar to the
case of hadroproduction shown in Fig.~6.

\section{Conclusion}

We have compared the $k_T$-factorization approach for heavy quark hadro-
and photopro\-duction at collider energies with the existing experimental
data. The agreement is reasonable when we take the cut-off (9) in Eqs.~(1)
and (2) while it does not practically depends on the value of QCD scale
$\mu$ in these equations. We present also predictions which can be used
as an additional test of our approach. Let us note that the
different expressions for the cut-off $\Delta$ result mainly into
normalization factor only, so our predictions from \cite{our1,our2}
obtained with the cut-off (6) can be used after renormalization with the
cut-off (9) too.

Another example of very successful agreement between experimental data and
$k_T$-factorization approach can be found in Ref.\cite{HKSST} based
on different assumptions.

We hope that future experiments will allow to distinguish between various
approaches. It is necessary to note that the theoretical results
concern the heavy quarks rather than experimentally observed hadron
production. The hadronization leads to several
important effects, however their description needs additional
phenomenological assump\-tions, see e.g. \cite{Likh,TNKN,Likh1,jdd,Shab}.

\subsection*{Acknowledgements}

We are grateful to M.G.Ryskin for a lot of very useful duscussions
and L.K.Gladilin for outlook the experimental situation at HERA.
This work was supported by grants NATO PSTCLG 977275 and RFBR
01-02-17095 and by the "Vedusch. Nauch. Shcool" RFFI 00-15-96610.

\newpage

\begin{center}
{\bf Figure Captions}
\end{center}

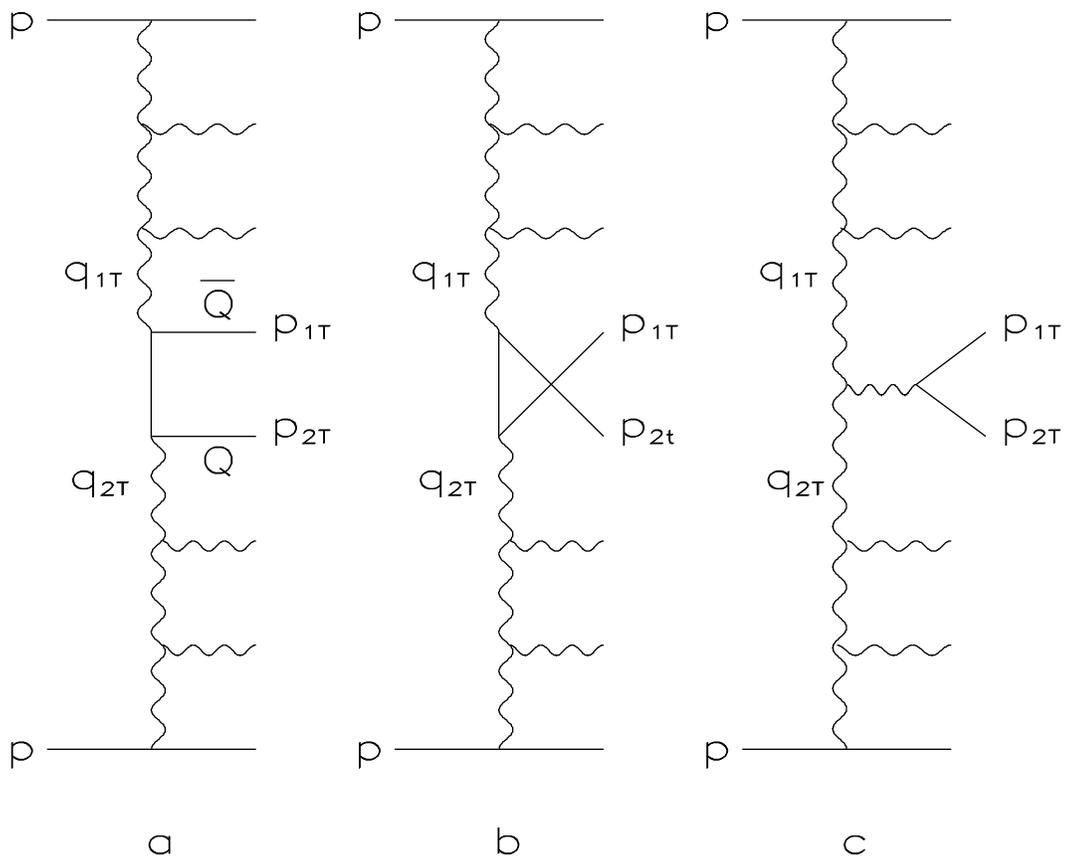
Fig.~1. Low order QCD diagrams for heavy quark production in $pp$
($p\overline{p}$) collisions via gluon-gluon fusion.

Fig.~2. The cross sections for beauty production with
$\vert y_1 \vert < 1$, in $p\bar{p}$ collisions at 1.8~TeV and their
description by the $k_T$-factorization approach with unintegrated gluon
distribution $f_g(x,q_T,\mu)$ given by Eq.~(1). The values of $\Delta$
in Eq.~(1), (2) are taken equal to $\mu/\sqrt(q_T^2 + \mu^2)$ (dotted
curves), to $\mu/\sqrt(q_t^2 + \mu^2 + q_t \mu )$ (solid curves), to
$\mu/(q_T + \mu)$ (dashed curves) and to $1 - q_T/\mu$ (dash-dotted
curves). The upper curves in the left-hand side of Fig.~2a correspond to
the scale value in Eqs.~(1) and (2) $\mu^2 = \hat{s}/4$, and the lower
curves to the scale value $\mu^2 = \hat{s}$. In the case of Fig.~2b
upper curves correspond to $\mu^2 = \hat{s}$ and lower ones to
$\mu^2 = \hat{s}/4$.

Fig.~3 The rapidity dependences of beauty production detected by decay
muons with $p^{\mu}_T > 5$ GeV (a) and $p^{\mu} > 8$ GeV (b) at 1.8 TeV.
The curve types are the same as in Fig.~2.

Fig.~4. Semi differential $b\bar{b}$ cross sections at
$\vert y_1 \vert < 1$, $\vert y_2 \vert < 1$, $p_{1T min}$ = 6.5~GeV
(a) and 8.75 GeV (b) and their description by the $k_T$-factorization
approach with unintegrated gluon distribution $f_g(x,q_T,\mu)$ given by
Eq.~(1). The curve types are the same as in Fig.~2.

Fig. ~5. The distributions of the total transverse momentum $p_{pair}$
for $b\bar{b}$ production in $p\bar{p}$ collisions at $\sqrt{s}$
= 1.8 TeV (a) and 14~TeV (b) and with the transverse momentum of one
heavy quark equal to 20~GeV, calculated with the unintegrated gluon
distribution $f_g(x,q_T,\mu)$ given by Eq.~(1). The curve types are the
same as in Fig.~2. The curves~1 are calculated with scale $\mu^2$ equal to
$\hat{s}$, and the curves~2 with the scale $\mu^2 = \hat{s}/4$.

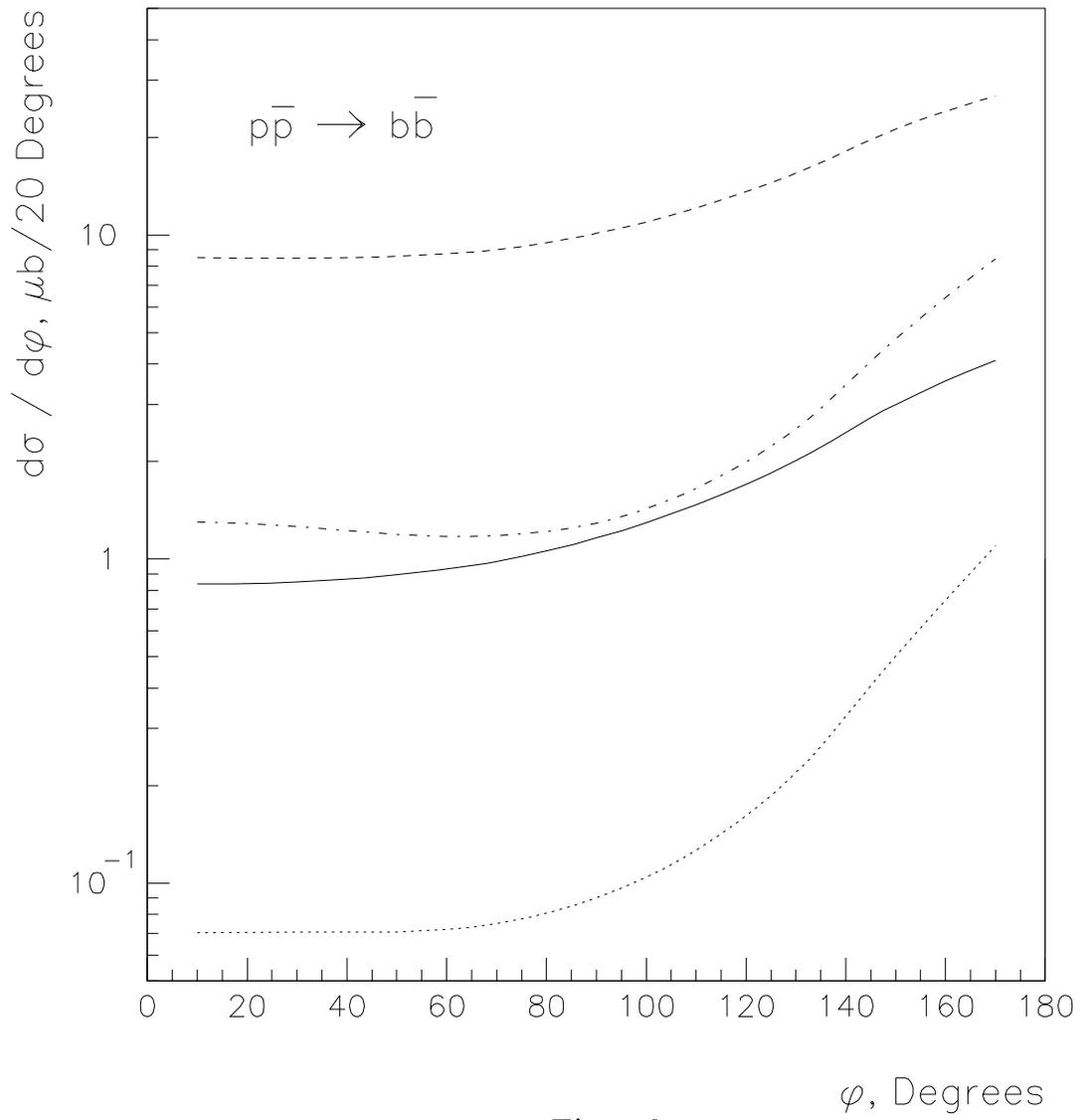
Fig.~6. The azimuthal angle distributions of $b$ quarks with
$\vert y_1^* \vert < 1$ and $\vert y_2^* \vert < 1$ produced in $pp$
($p\bar{p}$) collisions at 1.8~TeV (solid and dotted curves) and at
14~TeV (dashed and dash-dotted curves) for all $p_T$ of the produced quarks
(solid and dashed curves) and for both $p_{1T}, p_{2T} >$6~GeV (dotted and
dash-dotted curves).

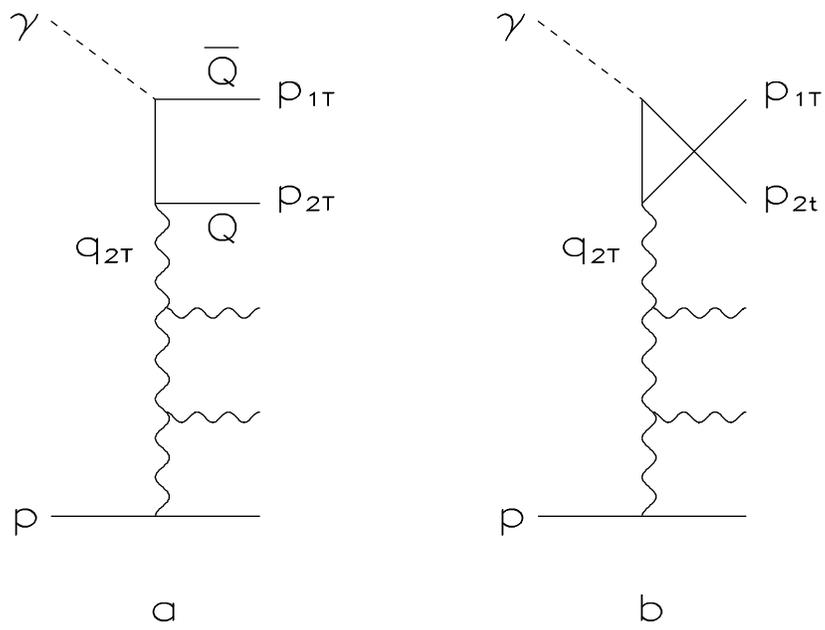
Fig.~7. Low order QCD diagrams for heavy quark production in direct
$\gamma p$ interactions via photon-gluon fusion.

Fig.~8. Total cross section of heavy quark photoproduction. The 
$\Delta$ values in Eq.~(1), (2) are taken equal to $\mu/\sqrt(q_T^2 + \mu^2)$
(dotted curves), to $\mu/\sqrt(q_t^2 + \mu^2 + q_t \mu )$ (solid
curves) and to $\mu/(q_T + \mu)$ (dashed curves). The curves 1 are
calculated with scale $\mu^2$ equal to $\hat{s}$, and the curves 2 with
scale $\mu^2 = \hat{s}/4$.

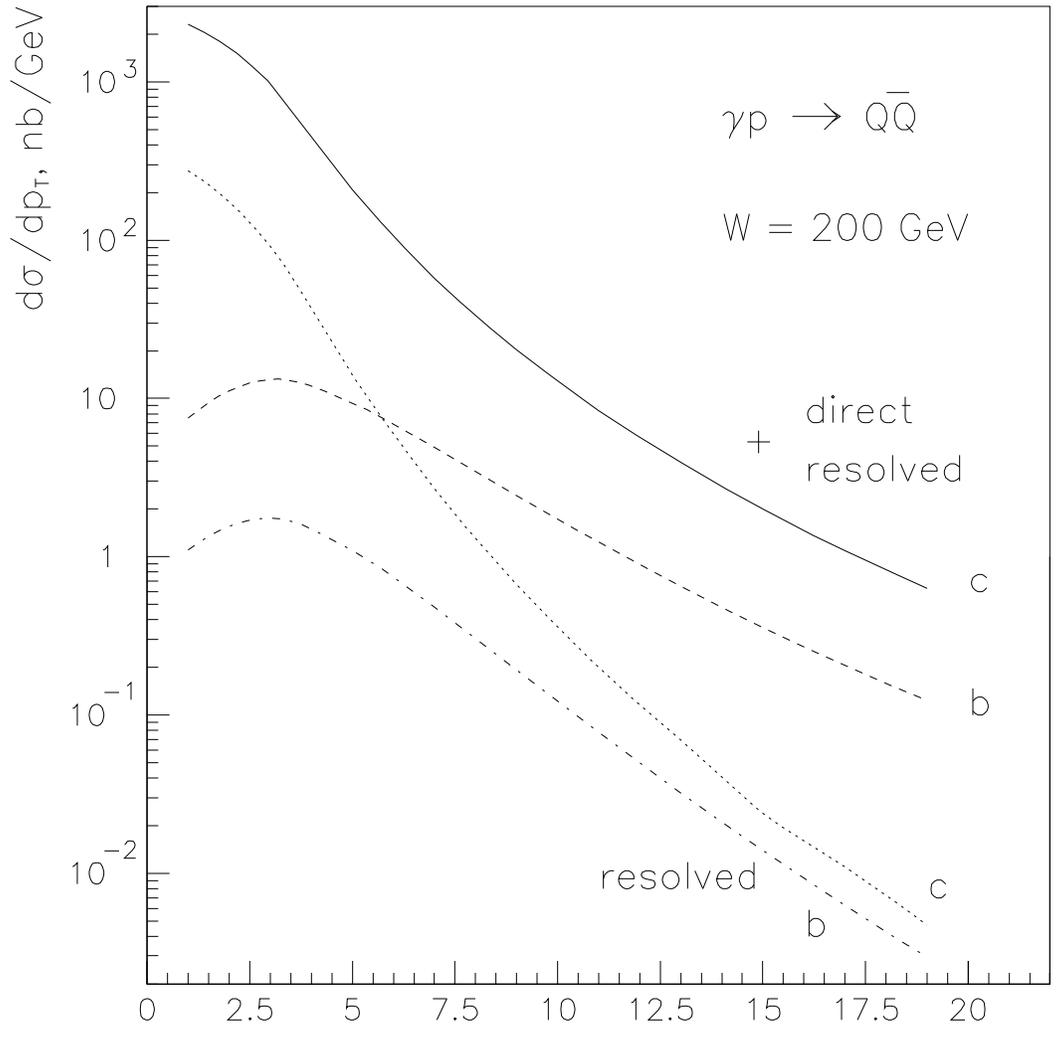
Fig.~9. The calculated $p_T$-distributions of charm and beauty
photoproduction at $W$ = 200~GeV (solid and dashed curves) with $\Delta$
in Eq.~(1), (2) equal to $\mu/\sqrt(q_T^2 + \mu^2 + q_T \mu )$ and
scale $\mu^2 = \hat{s}$. The resolved photon contributions are shown by
dotted and dash-dotted curves, respectively.

Fig.~10. The calculated rapidity distributions of charm (a) and beauty
(b) photopro\-duction at $W$ = 200~GeV for all $p_T$ (solid curve) and
$p_T > 6$~GeV (dashed curve) with $\Delta$ in Eq.~(1), (2) equal to
$\mu/\sqrt(q_T^2 + \mu^2 + q_T \mu )$ and scale $\mu^2 = \hat{s}$.
The resolved photon contributions for the presented two cases are shown
by dotted and dash-dotted curves, respectively.

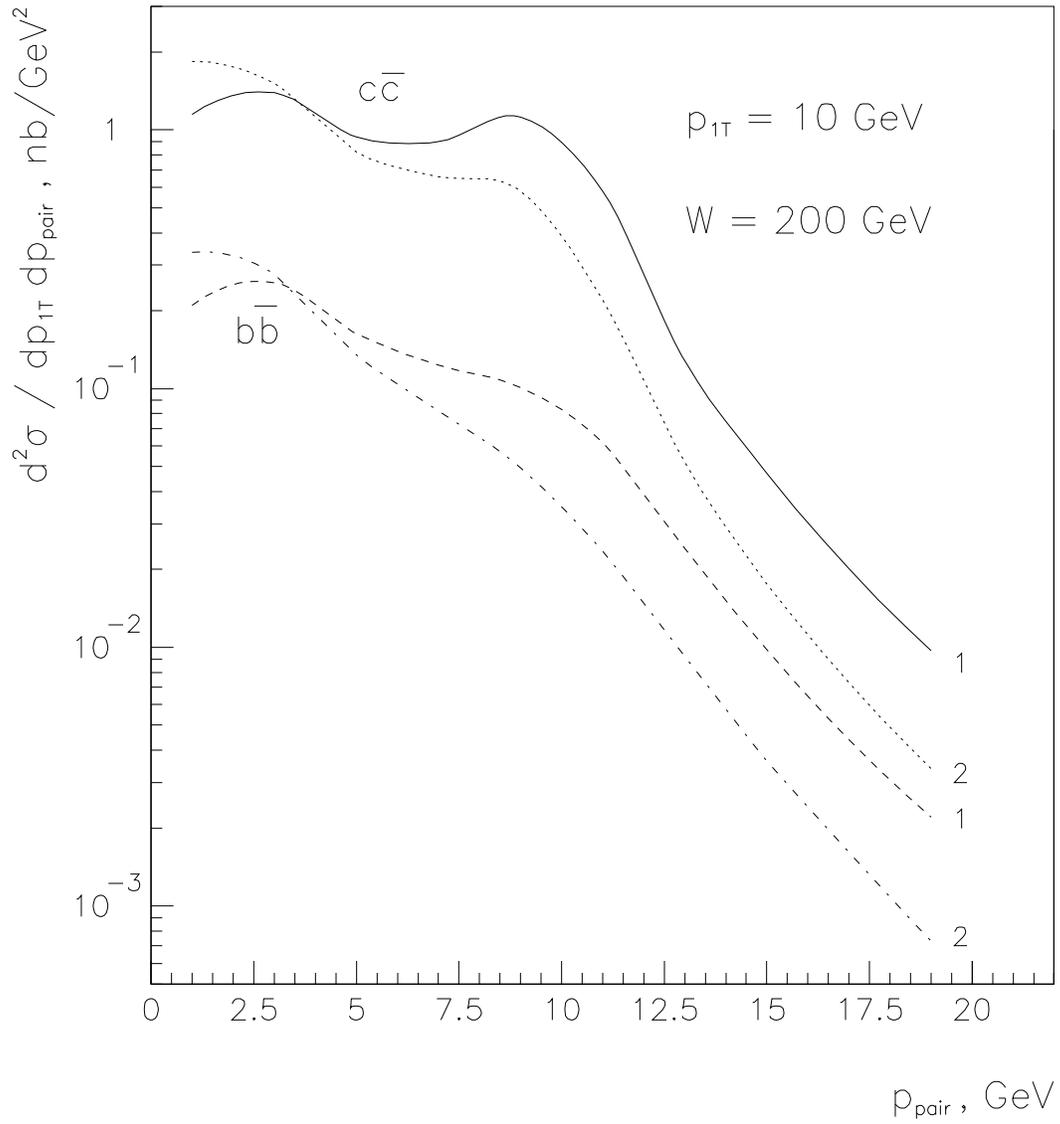
Fig.~11. The distributions of the total transverse momentum $p_{pair}$
for $c\bar{c}$ and $b\bar{b}$ produced in $\gamma p$ collisions at
$W$ = 200~GeV, and with the transverse momentum of a one heavy quark
equal to 10~GeV , calculated with the unintegrated gluon
distribution $f_g(x,q_T,\mu)$ given by Eq.~(1). The values of $\Delta$
in Eq.~(1), (2) are taken equal to $\mu/\sqrt(q_T^2 + \mu^2 + q_T \mu )$.
The curves 1 are calculated with scale $\mu^2$ equal to $\hat{s}$, and
the curves 2 with the scale $\mu^2 = \hat{s}/4$.

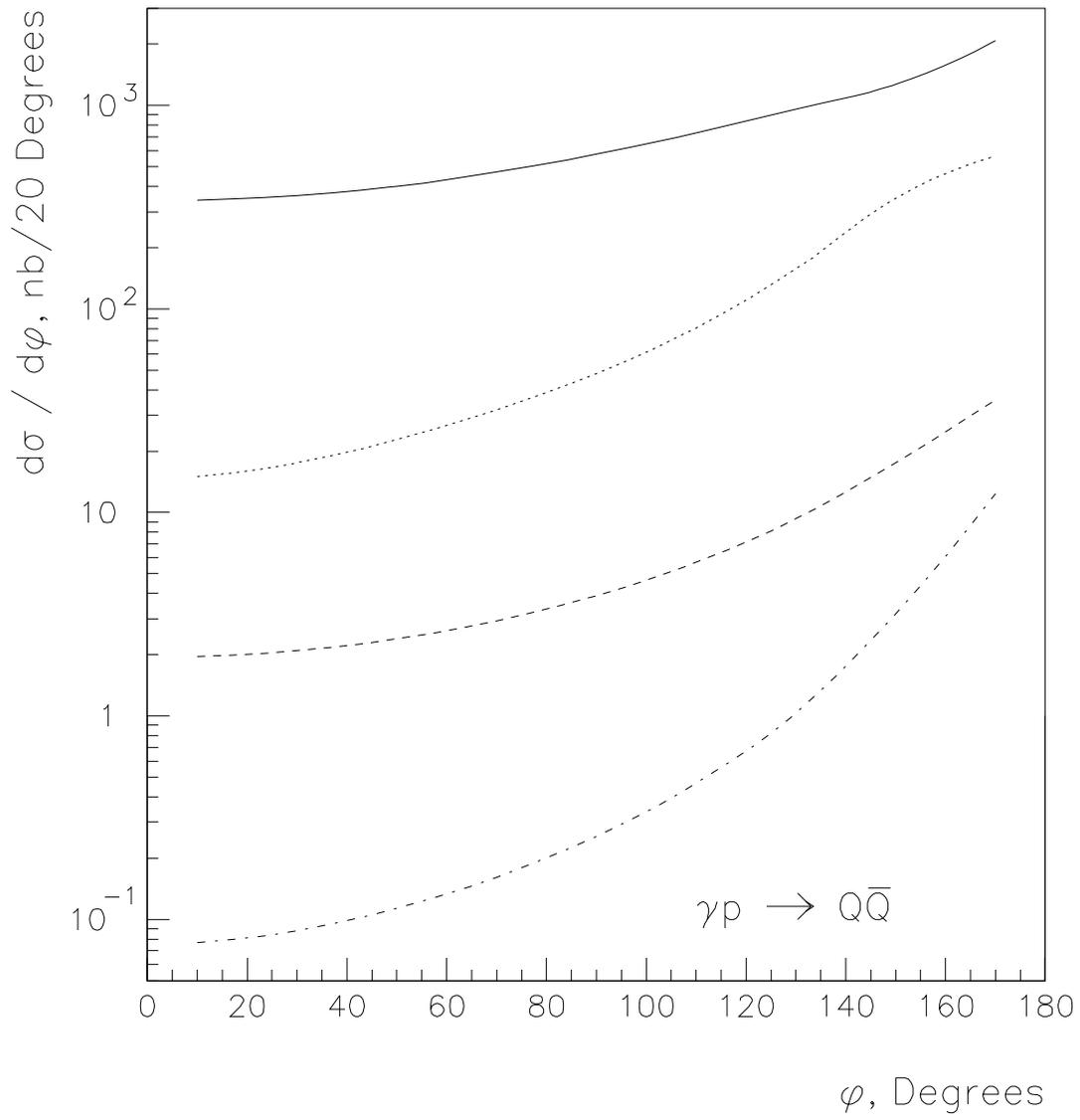
Fig.~12. The azimuthal angle distributions of $c$ and $b$ quarks with
all rapidities produced in $\gamma p$ collisions at 200~GeV for all
$p_T$ of produced quarks (solid and dashed curves, respectively), for
$p_{1T}, p_{2T} >$2~GeV for charm production (dotted curve) and for
$p_{1T}, p_{2T} >$5~GeV for beauty production (dash-dotted curves).

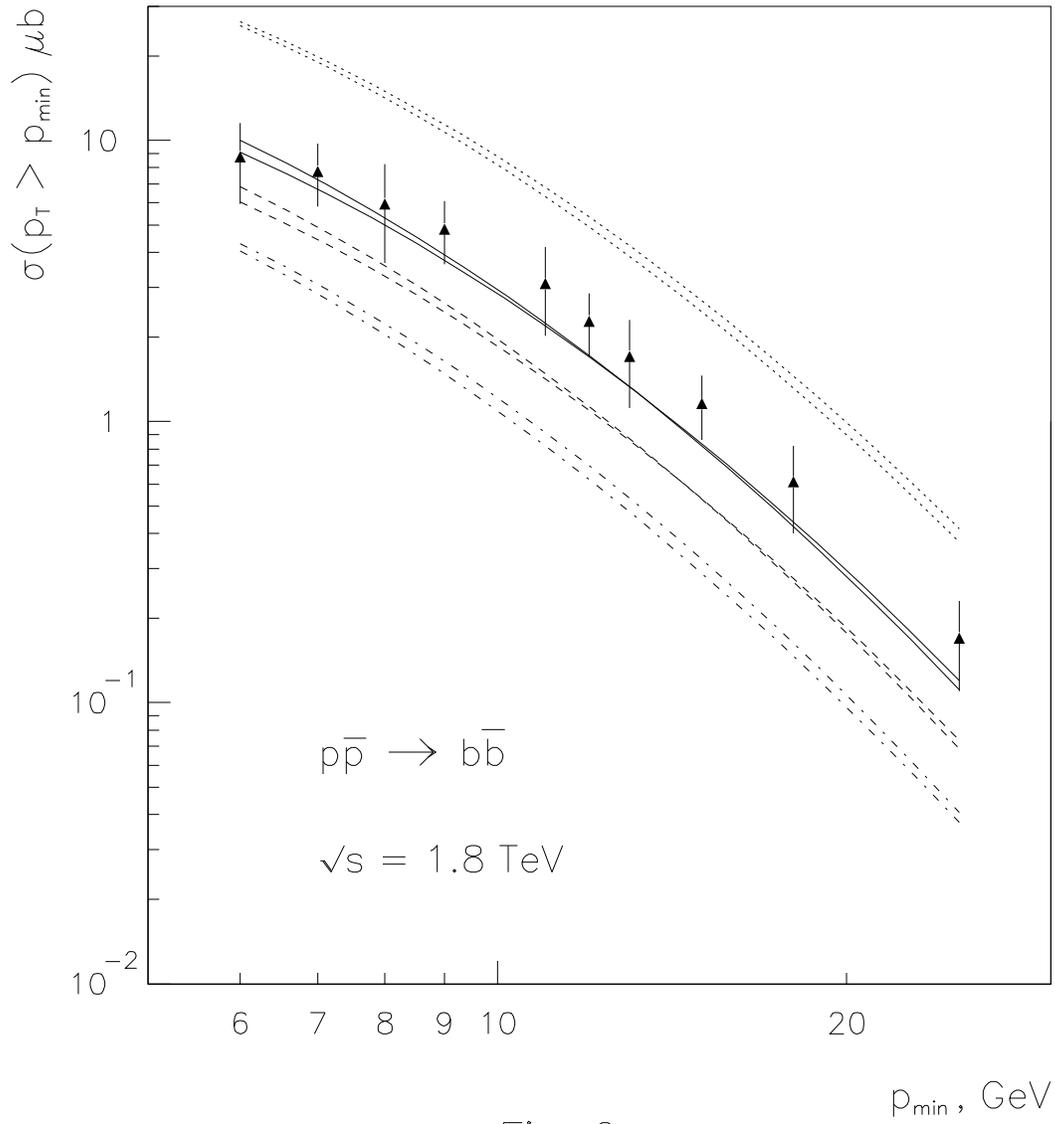
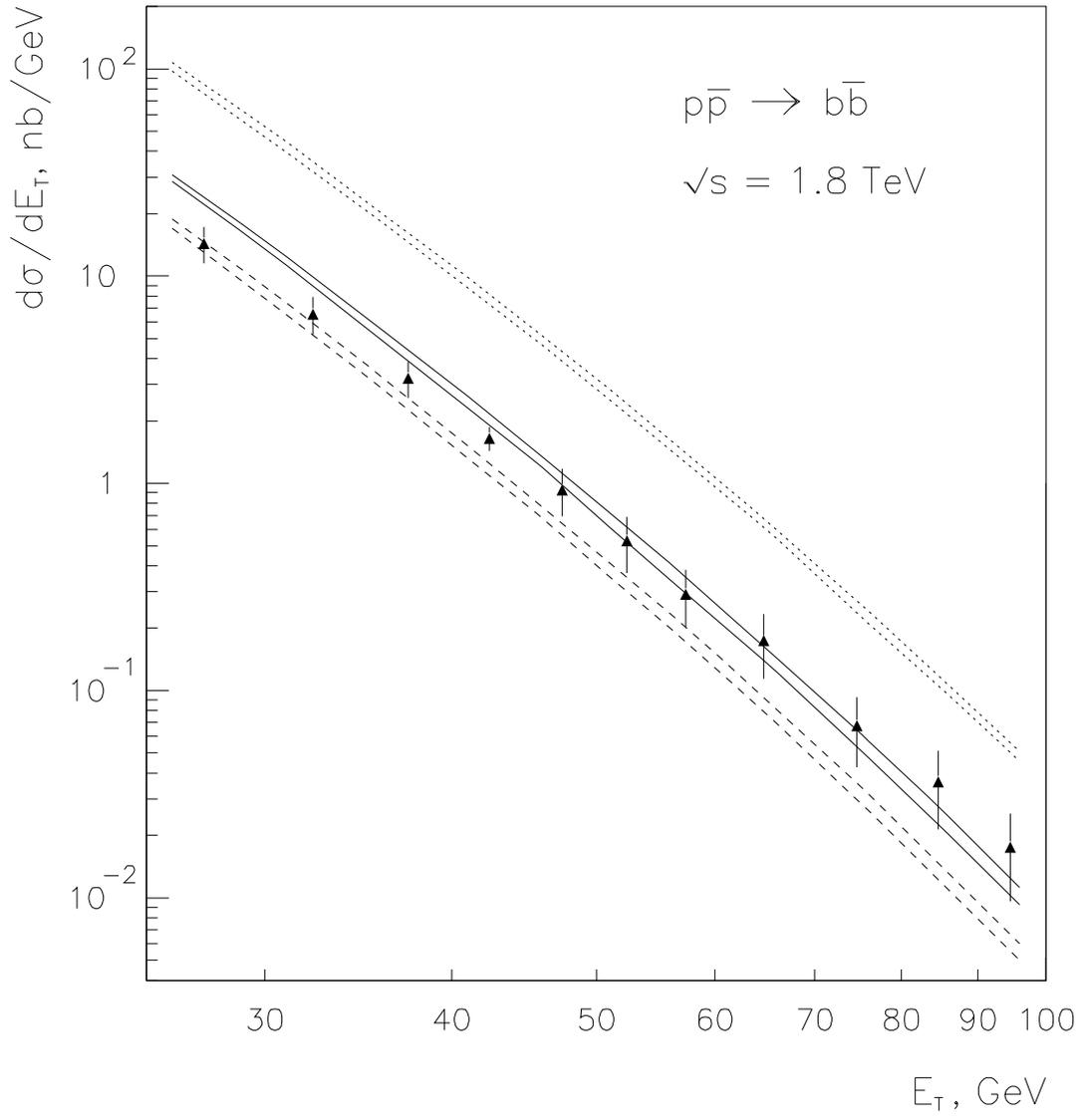
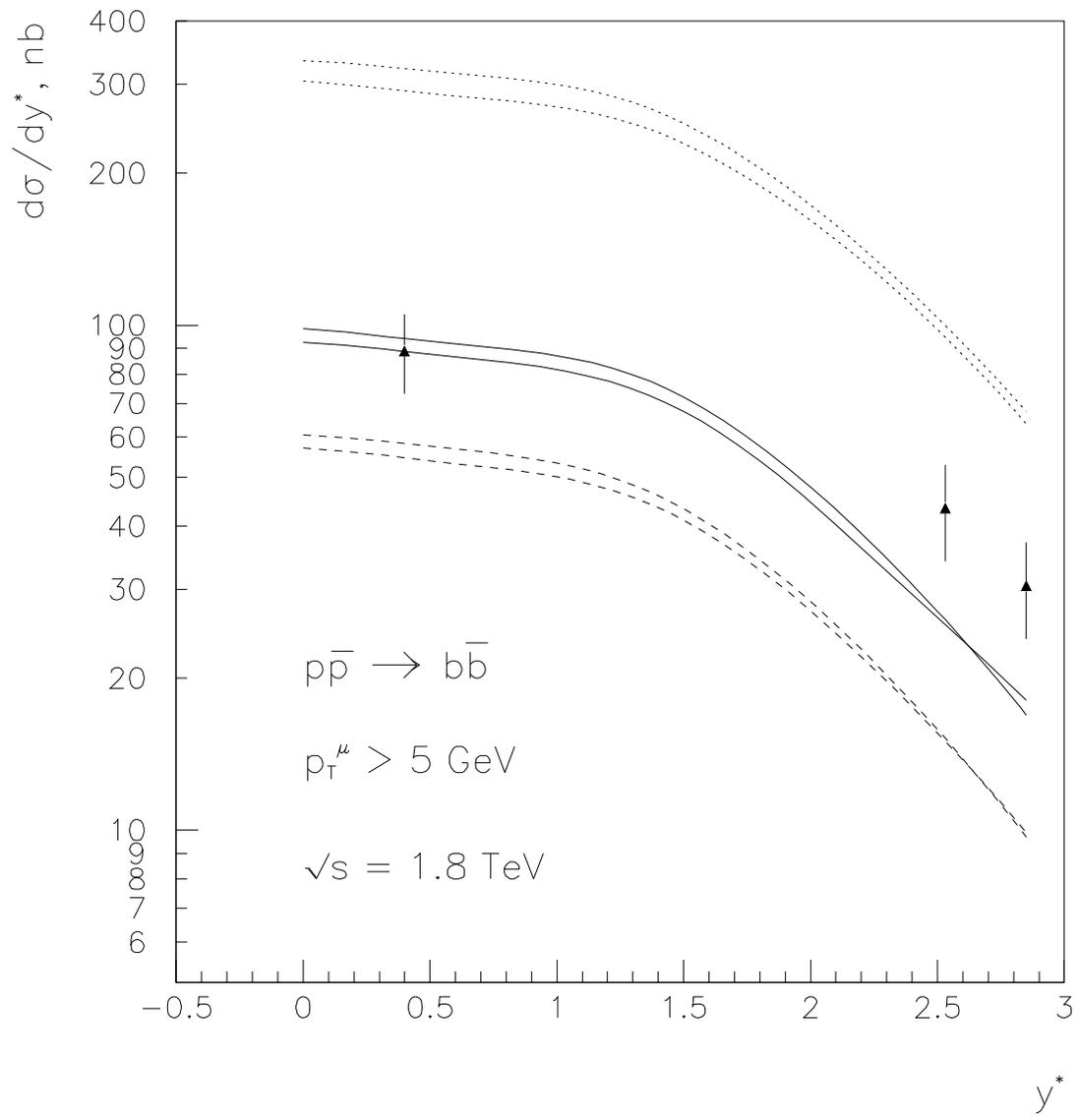
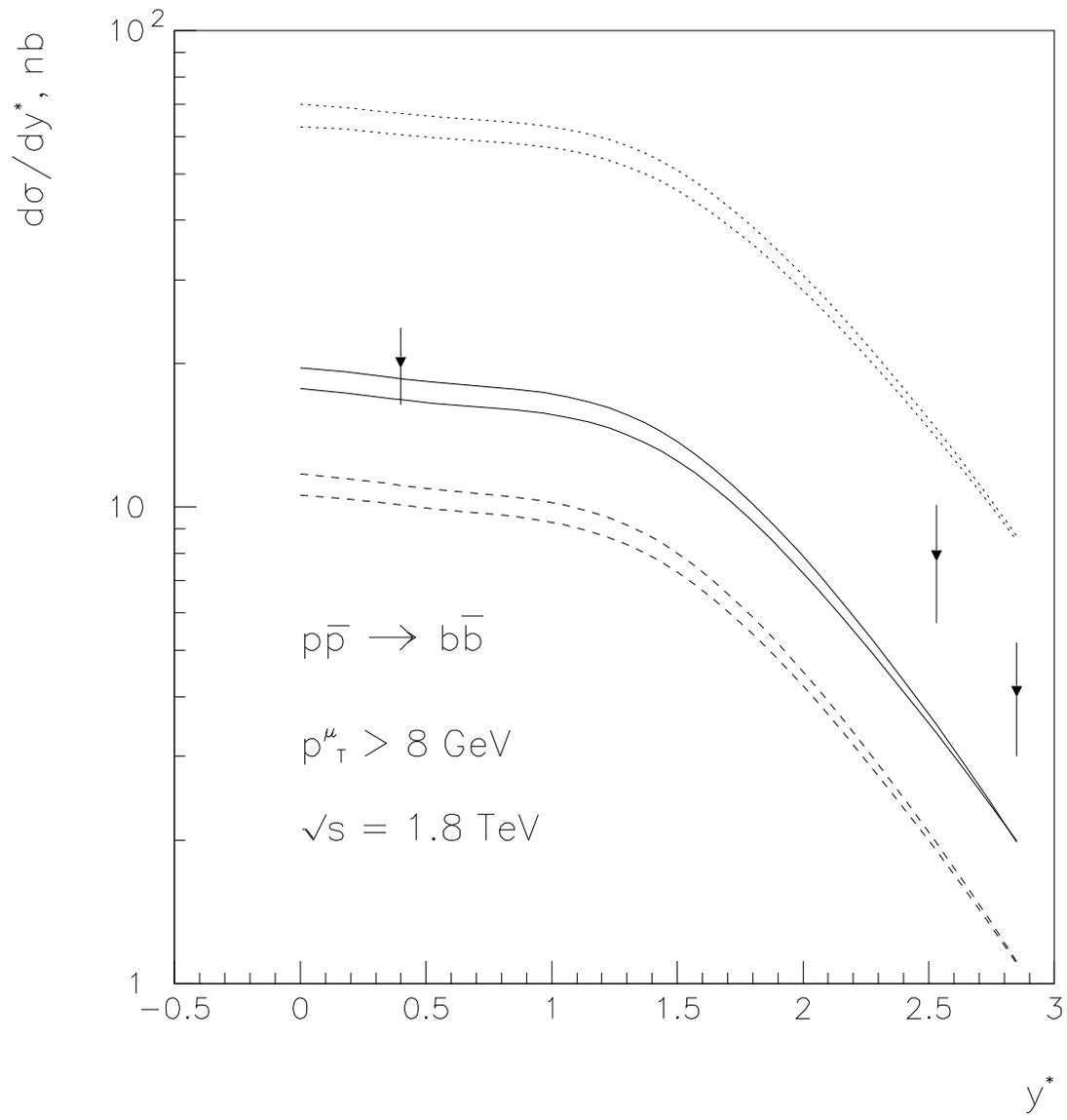
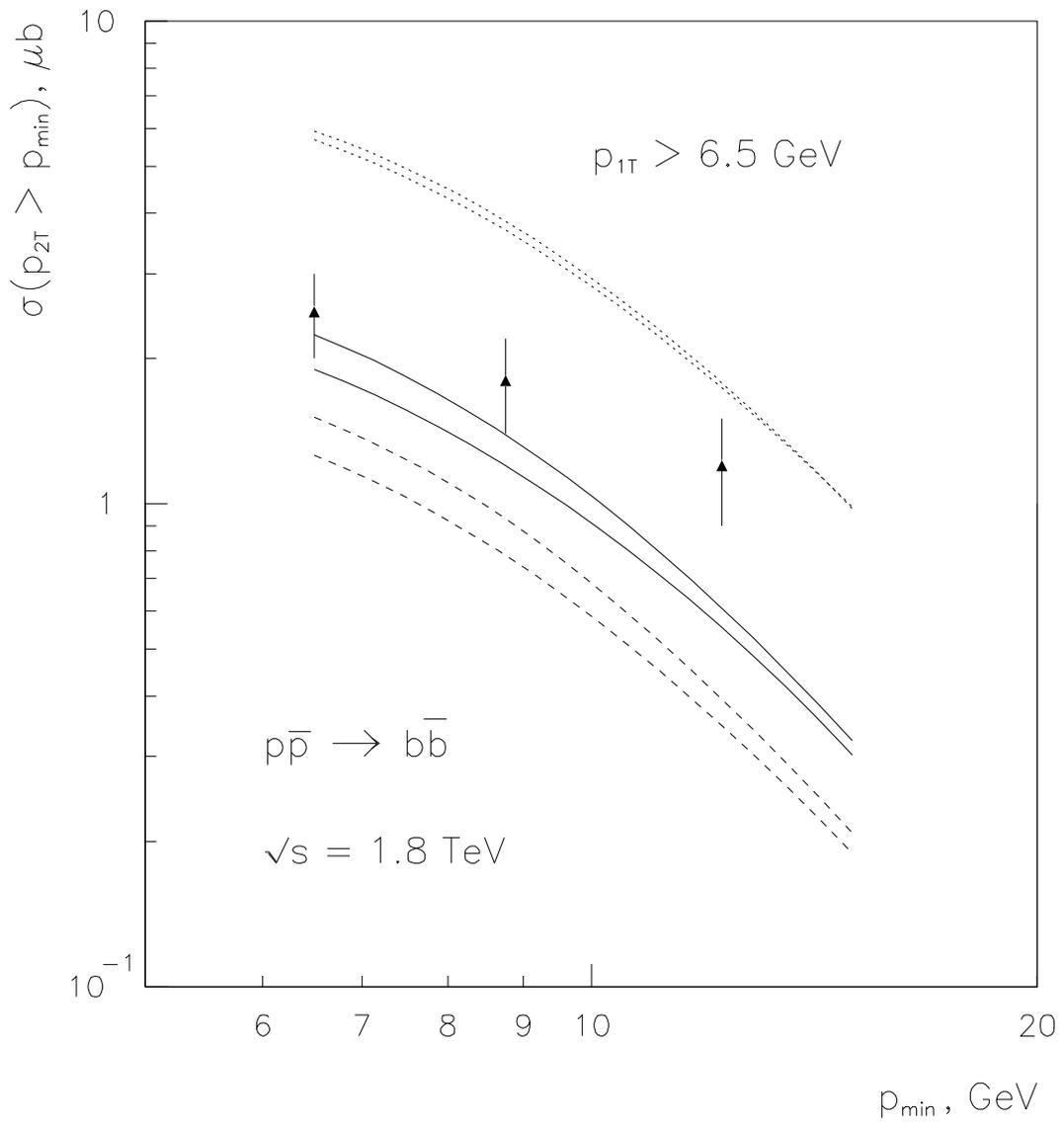
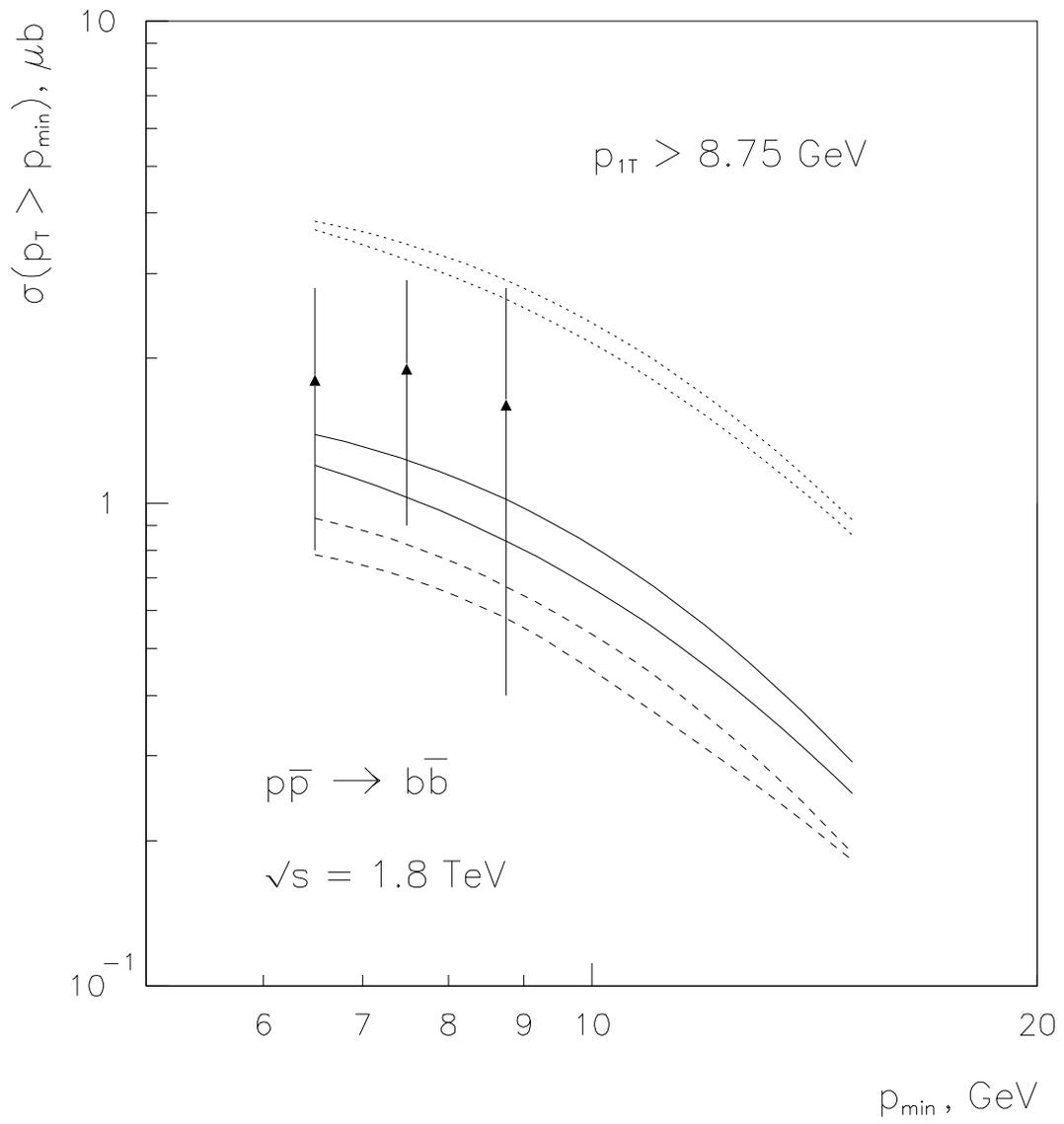
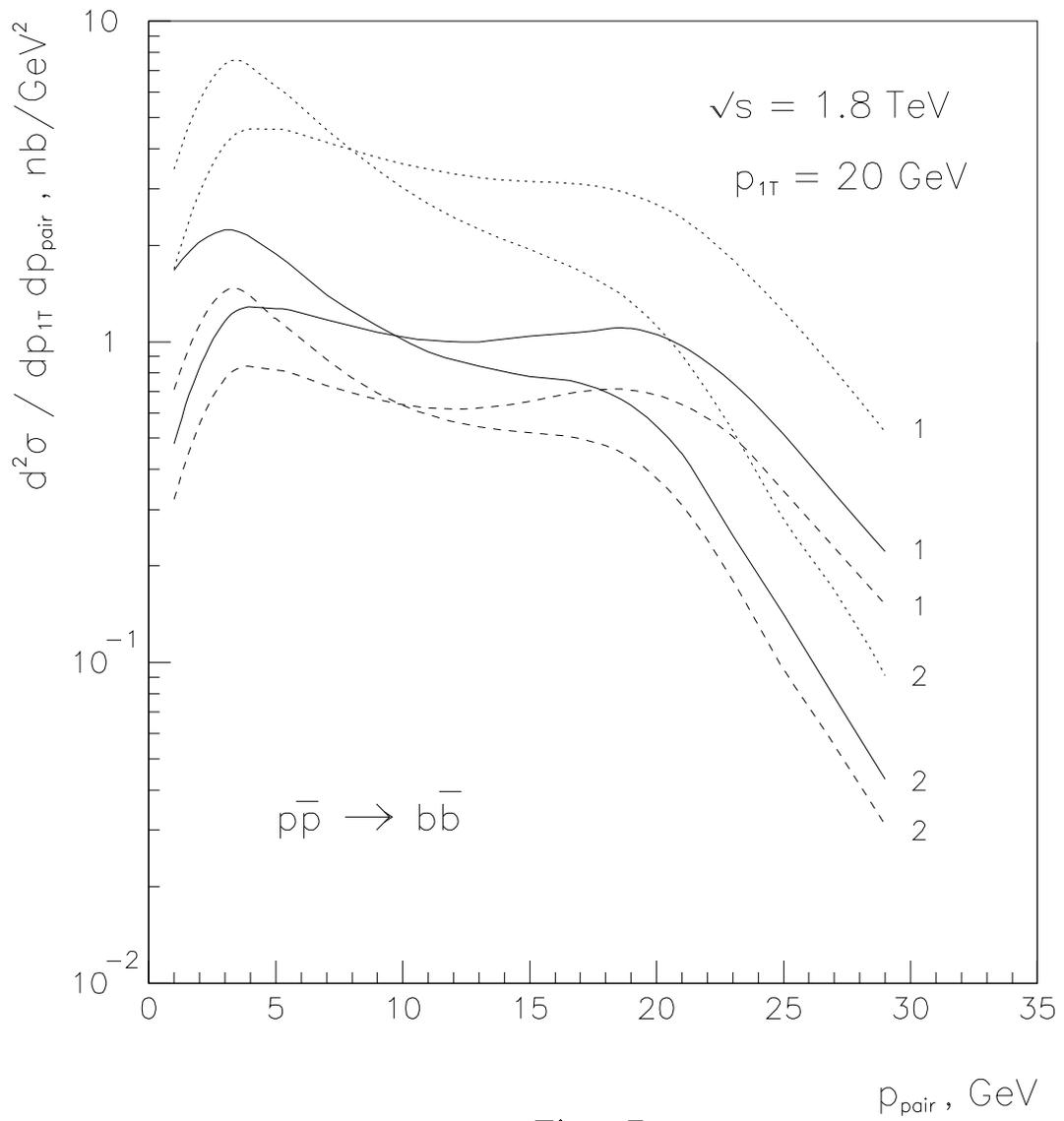
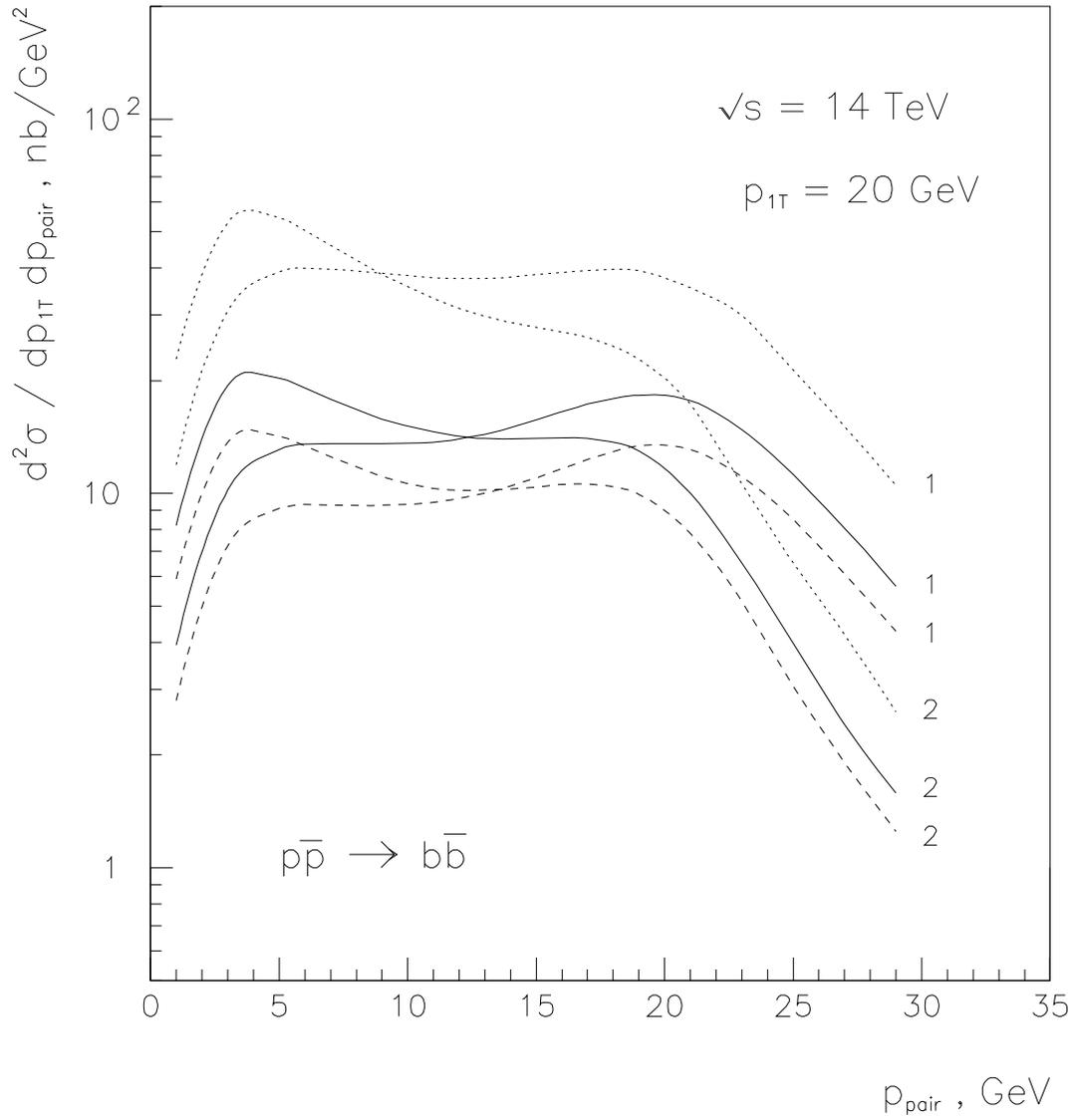
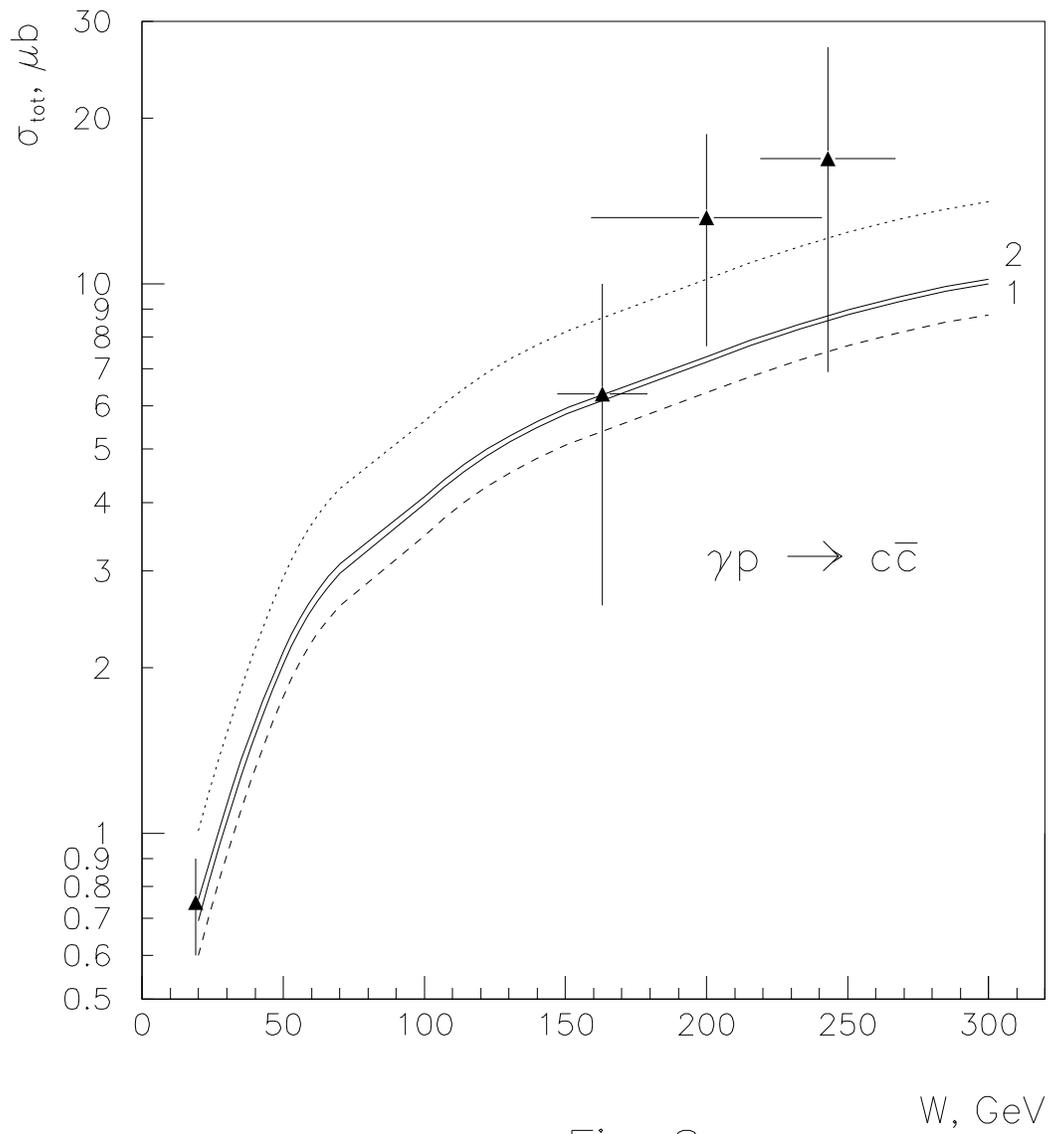
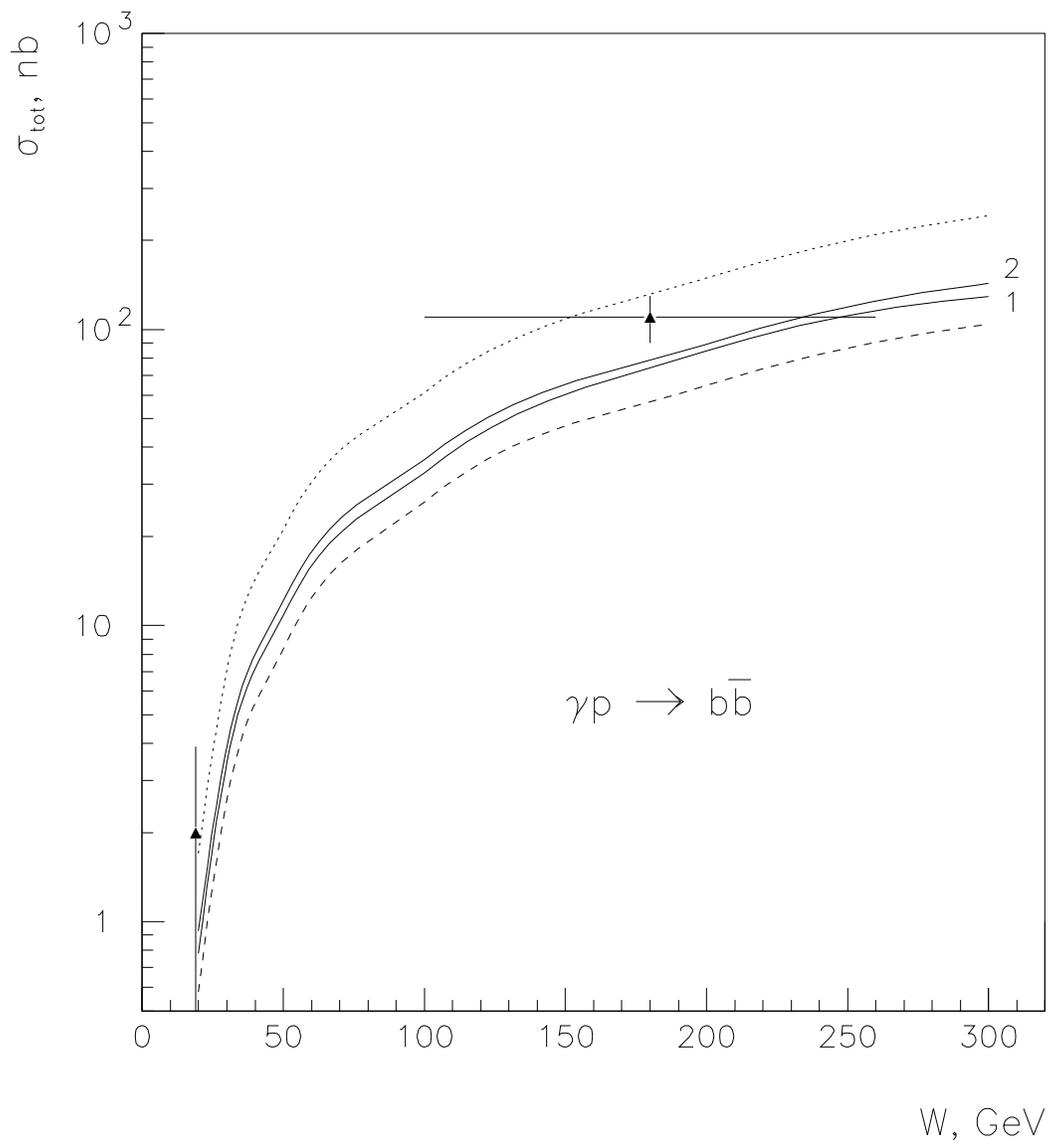
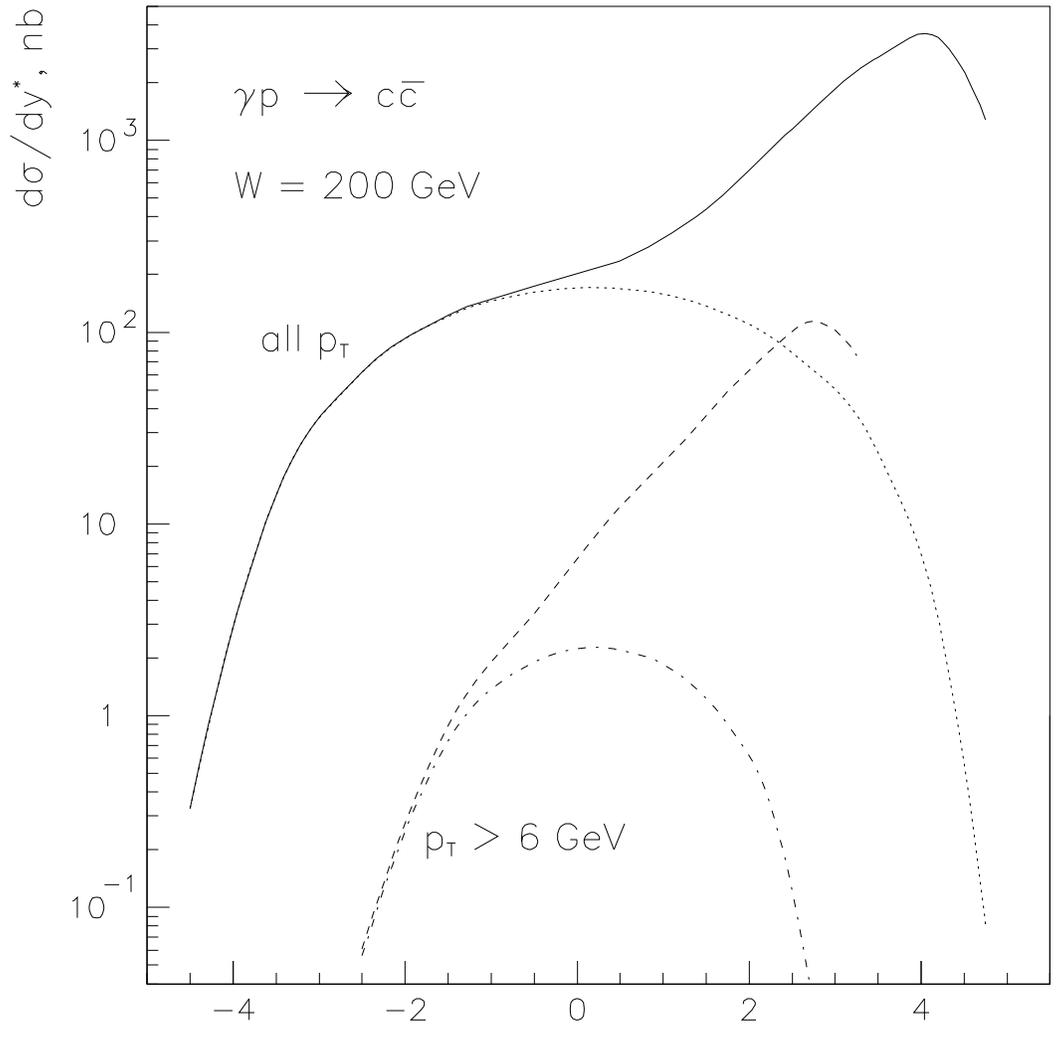
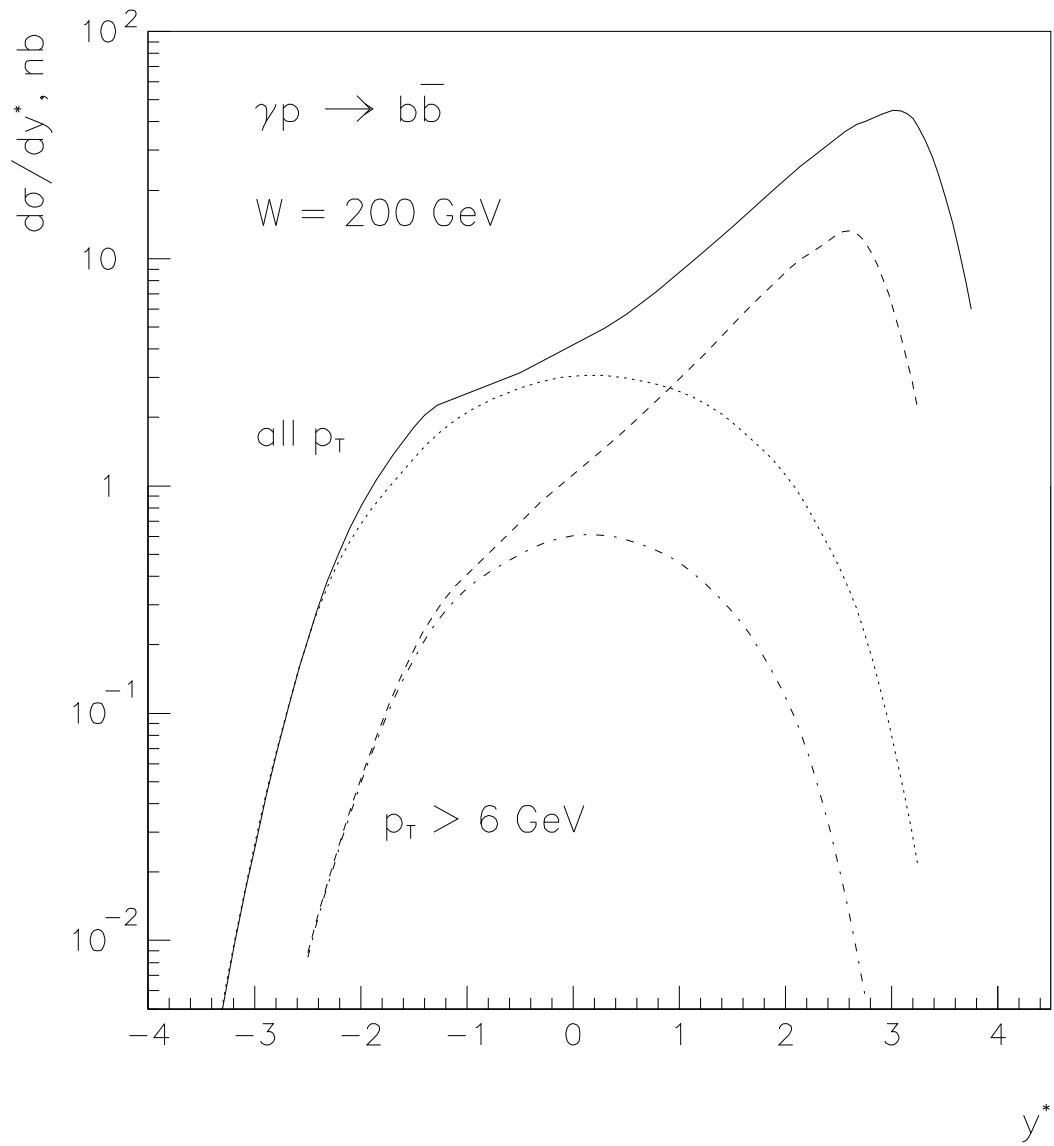
\end{document}